\documentclass[reprint,aps,pra,amsmath,superscriptaddress,amssymb,showpacs,footinbib,longbibliography,10pt]{revtex4-2}

\usepackage{graphicx}
\usepackage{color}

\usepackage{color}
\usepackage[colorlinks,bookmarks=false,citecolor=darkblue,linkcolor=red,urlcolor=blue]{hyperref}
\def\cred{
} 
\definecolor{darkred}{rgb}{0.7,0.0,0.0}
\definecolor{darkblue}{rgb}{0,0.02,0.45}

\newcommand{\kv}{\mathbf k}

\graphicspath{{figures/}}

\begin{document}

\title{MoP$_3$SiO$_{11}$: a $4d^3$ honeycomb antiferromagnet with disconnected octahedra}

\author{Danis I. Badrtdinov}
\affiliation{Theoretical Physics and Applied Mathematics Department, Ural Federal University, 620002 Yekaterinburg, Russia}

\author{Lei Ding}
\affiliation{ISIS Facility, Rutherford Appleton Laboratory, Harwell Oxford, Didcot, UK}
\affiliation{Institut N\'eel, CNRS and Universit\'e Joseph Fourier, 38042 Grenoble, France}

\author{Clemens Ritter}
\affiliation{Institut Laue-Langevin, BP 156, F-38042 Grenoble, France}

\author{Jan Hembacher}
\affiliation{Experimental Physics VI, Center for Electronic Correlations and Magnetism, Institute of Physics, University of Augsburg, 86135 Augsburg, Germany}

\author{Niyaz Ahmed}
\affiliation{Experimental Physics VI, Center for Electronic Correlations and Magnetism, Institute of Physics, University of Augsburg, 86135 Augsburg, Germany}
\affiliation{School of Physics, Indian Institute of Science
Education and Research Thiruvananthapuram-695551, India}

\author{Yurii Skourski}
\affiliation{Dresden High Magnetic Field Laboratory (HLD-EMFL), Helmholtz-Zentrum Dresden-Rossendorf, 01328 Dresden, Germany}

\author{Alexander A. Tsirlin}
\email{altsirlin@gmail.com}
\affiliation{Experimental Physics VI, Center for Electronic Correlations and Magnetism, Institute of Physics, University of Augsburg, 86135 Augsburg, Germany}
\affiliation{Theoretical Physics and Applied Mathematics Department, Ural Federal University, 620002 Yekaterinburg, Russia}

\begin{abstract}
We report the crystal structure and magnetic behavior of the $4d^3$ spin-$\frac32$ silicophosphate MoP$_3$SiO$_{11}$ studied by high-resolution synchrotron x-ray diffraction, neutron diffraction, thermodynamic measurements, and \textit{ab initio} band-structure calculations. Our data revise the crystallographic symmetry of this compound and establish its rhombohedral space group ($R\bar 3c$) along with the geometrically perfect honeycomb lattice of the Mo$^{3+}$ ions residing in disconnected MoO$_6$ octahedra. Long-range antiferromagnetic order with the propagation vector $\mathbf k=0$ observed below $T_N=6.8$\,K is a combined effect of the nearest-neighbor in-plane exchange coupling $J\simeq 2.6$\,K, easy-plane single-ion anisotropy $D\simeq 2.2 $\,K, and a weak interlayer coupling $J_c\simeq 0.8$\,K. The 12\% reduction in the ordered magnetic moment of the Mo$^{3+}$ ions and the magnon gap of $\Delta\simeq 7$\,K induced by the single-ion anisotropy further illustrate the impact of spin-orbit coupling on the magnetism. Our analysis puts forward single-ion anisotropy as an important ingredient of $4d^3$ honeycomb antiferromagnets despite their nominally quenched orbital moment. 
\end{abstract}

\maketitle


\section{Introduction}
\label{sec:introduction}
The honeycomb interaction geometry takes a special place in the physics of magnetic insulators. The bipartite nature of the honeycomb lattice excludes geometrical frustration for nearest-neighbor interactions~\cite{reger1989,weihong1991}, but allows interesting cases of exchange frustration in the presence of strong spin-orbit coupling when Kitaev and off-diagonal anisotropic terms become dominant interactions~\cite{rau2016,hermanns2018}. Experimental observations of Kitaev magnetism in $4d^5$ and $5d^5$ compounds with the effective spin-$\frac12$~\cite{winter2017}, such as $\alpha$-RuCl$_3$~\cite{takagi2019} and different polymorphs of Li$_2$IrO$_3$~\cite{tsirlin2021}, have triggered a broader interest in honeycomb magnets. On the theory side, proposals of Kitaev physics beyond the effective spin-$\frac12$ have been put forward~\cite{stavropoulos2019}, with implications for spin-orbit-coupled $d^4$ magnetic ions represented by Ru$^{4+}$~\cite{anisimov2019,chaloupka2019}. 

\begin{figure}
\includegraphics{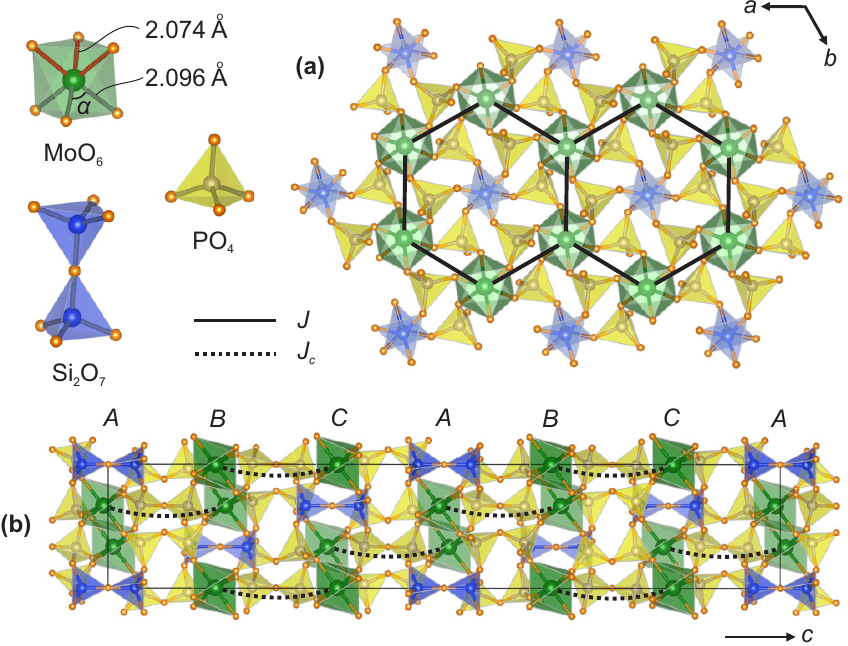}
\caption{Rhombohedral crystal structure of MoP$_3$SiO$_{11}$: (a) honeycomb planes of the MoO$_6$ octahedra separated by the PO$_4$ tetrahedra; the Si$_2$O$_7$ units center the hexagons; (b) $ABCABC$ stacking of the honeycomb planes and the interlayer couplings $J_c$ through the shortest Mo--Mo contacts between the planes. \texttt{VESTA} software was used for crystal structure visualization~\cite{vesta}.
}
\label{fig:Crystal}
\end{figure}

Compared to $d^5$ and $d^4$, the $4d^3$ case of Ru$^{5+}$ may seem less exotic, because the half-filling of the $t_{2g}$ shell quenches the orbital moment. Nevertheless, the Affleck-Kennedy-Lieb-Tasaki (AKLT) phase, a model entangled state for quantum computation~\cite{wei2011}, was predicted to appear in $d^3$ systems in the limit of weak Hund's coupling $J_H$~\cite{janusz2015,jakab2016}. On increasing $J_H$, this AKLT phase transforms into a N\'eel-ordered state that has been observed  experimentally in SrRu$_2$O$_6$~\cite{hiley2014} and caused significant attention because of its very high N\'eel temperature of 565\,K and an unusually low ordered moment of only $1.3-1.4$\,$\mu_B$~\cite{hiley2015,tian2015} compared to 3\,$\mu_B$ expected for a spin-$\frac32$ ion. These observations could not be explained on the level of a simple nearest-neighbor Heisenberg Hamiltonian. An unusual electronic state with hexagon molecular orbitals was proposed~\cite{streltsov2015} and subsequently investigated theoretically and experimentally~\cite{pchelkina2016,okamoto2017,hariki2017,ponosov2019}, although a more conventional description on the level of a Heisenberg Hamiltonian with an additional single-ion anisotropy term~\cite{singh2015} proved sufficient for explaining magnetic excitations of this material~\cite{suzuki2019}. The isoelectronic compound AgRuO$_3$ shows a similar phenomenology, albeit with a somewhat lower $T_N$ of 342\,K~\cite{prasad2017,schnelle2021}.

Beyond ruthenates, several recent studies discussed the possibility of Kitaev interactions and other anisotropic exchange interactions in spin-$\frac32$ honeycomb ferromagnets, such as CrI$_3$~\cite{xu2018,xu2020,lee2020,chen2020,stavropoulos2021}. This raises the question whether the on-site (single-ion) anisotropy or different inter-site effects (Kitaev anisotropy, hexagon molecular orbitals) should be used to describe magnetism of the $d^3$ honeycomb systems. To address this question, we consider MoP$_3$SiO$_{11}$ silicophosphate~\cite{leclaire1987}, the $4d^3$ honeycomb antiferromagnet with disconnected transition-metal octahedra (Fig.~\ref{fig:Crystal}). Increased separations between the magnetic ions suppress intersite effects and expose single-ion anisotropy as the dominant anisotropy term despite the nominally quenched orbital moment. 


\section{Methods}

Polycrystalline samples of MoP$_3$SiO$_{11}$ were prepared by a two-step solid-state reaction. First, a mixture of MoO$_3$, SiO$_2$, and (NH$_4)_2$HPO$_4$ taken in the 0.5:1:3 molar ratio was annealed in air at 600\,$^{\circ}$C for 24 hours. The reaction produced a dark-green melted product that was re-ground, mixed with Mo powder (Alfa Aesar, 2-4 micron particle size) according to the MoP$_3$SiO$_{11}$ stoichiometry, and annealed at 870\,$^{\circ}$C for 100 hours. This second annealing was performed in a sealed quartz tube filled with 300\,mbar of argon to prevent oxidation of Mo$^{3+}$. The brownish-green product {\cred was phase-pure MoP$_3$SiO$_{11}$ when smaller samples with the total mass of $0.1-0.2$\,g were prepared. For larger samples, minor amounts of the MoP$_2$O$_7$ impurity were observed.}

High-resolution x-ray diffraction (XRD) data~\cite{esrf} were collected at room temperature at the ID22 beamline of the European Synchrotron Radiation Facility (ESRF), Grenoble using the wavelength of 0.35424\,\r A. The sample was loaded into a thin borosilicate glass capillary and spun during the measurement. The diffracted signal was measured by 9 scintillation detectors, each preceded by a Si (111) analyzer crystal. 

Neutron diffraction data~\cite{neutron} were collected at the D2B ($\lambda=1.594$\,\r A) and D20 ($\lambda=2.41$\,\r A) instruments at the Institut Laue-Langevin (ILL), Grenoble. The powder sample of MoP$_3$SiO$_{11}$ was loaded into a vanadium container and cooled down to 1.5\,K with the standard Orange cryostat. Rietveld refinements were performed in \texttt{Jana2006}~\cite{jana2006} and \texttt{Fullprof}~\cite{fullprof}.

Temperature-dependent magnetic susceptibility was measured with MPMS3 SQUID magnetometer from Quantum Design in the temperature range of $1.8-300$\,K in applied fields up to 7\,T. Magnetization measurements up to 56\,T were performed in Dresden High Magnetic Field Laboratory at 1.4\,K on a powder sample loaded into a thin kapton tube. Heat capacity was measured in the temperature range of $1.8-300$\,K in magnetic fields up to 14\,T on a pressed pellet by the relaxation method using Quantum Design PPMS.

Electronic structure calculations were performed on the level of density-functional theory (DFT) utilizing the generalized gradient approximation (GGA)~\cite{pbe96}. To this end, Quantum Espresso~\cite{espresso} and Vienna ab initio Simulation Package (VASP)~\cite{vasp1,vasp2} codes with the plane-wave basis set were used. The energy cutoff was set at 700\,eV and the energy convergence criteria was 10$^{-6}$\,eV. For the Brillouin-zone integration, a 4$\times$4$\times$4 Monkhorst-Pack mesh was used.  

Thermodynamic properties of the magnetic model parameterized by DFT were obtained from quantum Monte-Carlo (QMC) simulations performed using the \texttt{loop}~\cite{loop} and \verb|dirloop_sse|~\cite{dirloop} algorithms of the \texttt{ALPS} simulation package~\cite{alps}. We performed simulations on $L \times L \times L/2$  finite lattices with $L\leq 20$, where the unit cell contains eight magnetic sites with spin $S=3/2$. 


\section{Experimental results}
\label{sec:experimental data}

\subsection{Crystal structure}

\begin{figure}
\includegraphics{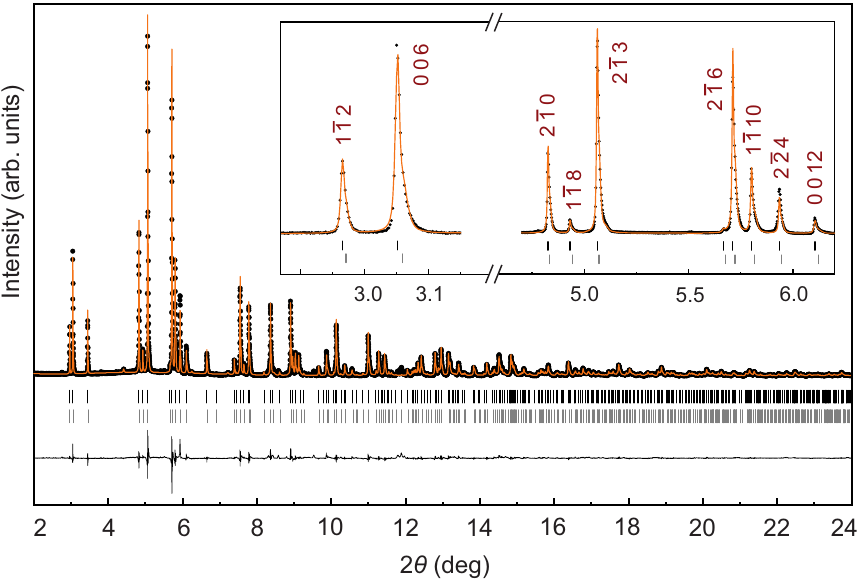}
\caption{\label{fig:xrd}
Rietveld refinement for the room-temperature high-resolution XRD data {\cred collected on the phase-pure sample of MoP$_3$SiO$_{11}$. The refined model included} two $R\bar 3c$ phases with the slightly different $c$ lattice parameters [$c_1=39.9085(5)$\,\r A, $c_2=39.812(2)$\,\r A]. The inset shows the asymmetric peak broadening, which is more pronounced in the $hkl$ reflections with large $l$. The refinement residuals are $R_I=0.045$ and $R_p=0.072$.
}
\end{figure}

The crystal structure of MoP$_3$SiO$_{11}$ (Fig.~\ref{fig:Crystal}) features honeycomb layers of the MoO$_6$ octahedra, which are separated by the PO$_4$ tetrahedra. This type of structural geometry is very different from the dense layers of transition-metal octahedra typically encountered in hexagonal magnets, such as $\alpha$-Li$_2$IrO$_3$ and SrRu$_2$O$_6$. The increased nearest-neighbor Mo--Mo distance of 4.9\,\r A allows the hexagons to accommodate large Si$_2$O$_7$ pyrosilicate units in the center, whereas the PO$_4$ tetrahedra located above and below these hexagons condense into P$_2$O$_7$ units that connect adjacent layers stacked along the $c$ axis. 

Although monoclinic $C2/c$ symmetry has been reported for MoP$_3$SiO$_{11}$ earlier~\cite{leclaire1987}, several structural parameters indicate that this crystal structure can be described as rhombohedral. Indeed, the $b/a$ ratio of 1.7329(4) is very close to $\sqrt 3$, whereas MoO$_6$ octahedra are three-fold symmetric. Moreover, the trigonal $R\bar 3c$ symmetry was reported for a sister compound RuP$_3$SiO$_{11}$~\cite{fukuoka1996}. 

\begin{table}
\caption{\label{tab:structure}
Atomic positions and atomic displacement parameters ($U_{\rm iso}$, in $10^{-2}$\,\r A$^2$) for MoP$_3$SiO$_{11}$ refined against the D2B data at 1.5\,K and 295\,K. The lattice parameters are $a=8.3952(3)$\,\r A, $c=39.869(2)$\,\r A at 1.5\,K and $a=8.4015(3)$\,\r A, $c=39.847(2)$\,\r A at 295\,K. The space group is $R\bar 3c$. The refinement residuals are $R_I=0.025$, $R_p=0.023$ at 1.5\,K and $R_I=0.033$, $R_p=0.025$ at 295\,K. The $U_{\rm iso}$ for oxygen atoms have been constrained. The cif-files are available as Supplemental Material~\cite{suppl}.
}
\begin{ruledtabular}
\begin{tabular}{cc@{\hspace{1em}}c@{\hspace{1em}}cccc}
      & Site &     & $x/a$       & $y/b$     & $z/c$     & $U_{\rm iso}$ \\\hline
 Mo & $12c$ & 1.5\,K & 0         & 0         & 0.1592(1) & 0.05(9) \\\smallskip
    &       & 295\,K & 0         & 0         & 0.1593(1) & 0.33(11) \\
 P  & $36f$ & 1.5\,K & 0.3712(4) & 0.0331(4) & 0.1193(1) & 0.21(8) \\\smallskip
    &       & 295\,K & 0.3702(5) & 0.0323(5) & 0.1190(1) & 0.64(9) \\
 Si & $12c$ & 1.5\,K & 0         & 0         & 0.4602(2) & 0.9(2)  \\\smallskip
    &       & 295\,K & 0         & 0         & 0.4607(2) & 0.9(2)  \\
 O1 & $36f$ & 1.5\,K & 0.2868(4) & 0.8178(4) & 0.1132(1) & 0.23(3) \\\smallskip
    &       & 295\,K & 0.2879(5) & 0.8180(5) & 0.1134(1) & 0.89(5) \\
 O2 & $36f$ & 1.5\,K & 0.2225(5) & 0.0751(5) & 0.1270(1) & 0.23(5) \\\smallskip				
    &       & 295\,K & 0.2237(6) & 0.0750(6) & 0.1271(1) & 0.89(5) \\
 O3 & $36f$ & 1.5\,K & 0.5279(4) & 0.1025(4) & 0.1435(1) & 0.23(5) \\\smallskip
    &       & 295\,K & 0.5282(5) & 0.1048(5) & 0.1434(1) & 0.89(5) \\
 O4 & $18e$ & 1.5\,K & 0.7867(5) & 0         & $\frac14$ & 0.23(5) \\\smallskip
    &       & 295\,K & 0.7858(6) & 0         & $\frac14$ & 0.89(5) \\
 O5 & $6a$  & 1.5\,K & 0         & 0         & 0         & 0.23(5) \\
    &       & 295\,K & 0         & 0         & 0         & 0.89(5) \\
\end{tabular}
\end{ruledtabular}
\end{table}

We used high-resolution XRD to verify the rhombohedral symmetry of MoP$_3$SiO$_{11}$. No peak splitting could be observed, and the data were successfully refined in the rhombohedral structure ($R\bar 3c$). A closer inspection of the XRD pattern reveals that some of the peaks are asymmetrically broadened (Fig.~\ref{fig:xrd}), but this broadening is incompatible with the monoclinic distortion. For example, the broadening is observed for the $006$ reflection that would not be split in $C2/c$. The asymmetry is most pronounced in the $hkl$ reflections with large $l$ and can be described by a second phase with the reduced lattice parameter $c$. This rather subtle broadening effect is not detectable using lab XRD and may be related to the stacking disorder. It does not affect any of the magnetic properties shown below. 

Crystal structure refinements of the neutron diffraction data (Table~\ref{tab:structure}) were also performed in $R\bar 3c$, but with only one phase because the asymmetric broadening was beyond the resolution of the D2B diffractometer. The refinement confirms the regular hexagonal arrangement of the Mo$^{3+}$ ions located in trigonally distorted MoO$_6$ octahedra with the Mo--O distances of 2.074(5) and 2.096(7)\,\r A, respectively. The stacking of the honeycomb planes follows the $ABCABC$ sequence with six layers per unit cell (Fig.~\ref{fig:Crystal}b). No significant structural changes are observed upon cooling from room temperature to 1.5\,K. The shortest interlayer Mo--Mo separation of 7.264\,\r A is much longer than that in other honeycomb magnets, such as SrRu$_2$O$_6$ (5.23\,\r A~\cite{hiley2014}) and CrI$_3$ (6.59\,\r A~\cite{mcguire2015}). 


\begin{figure}
\includegraphics{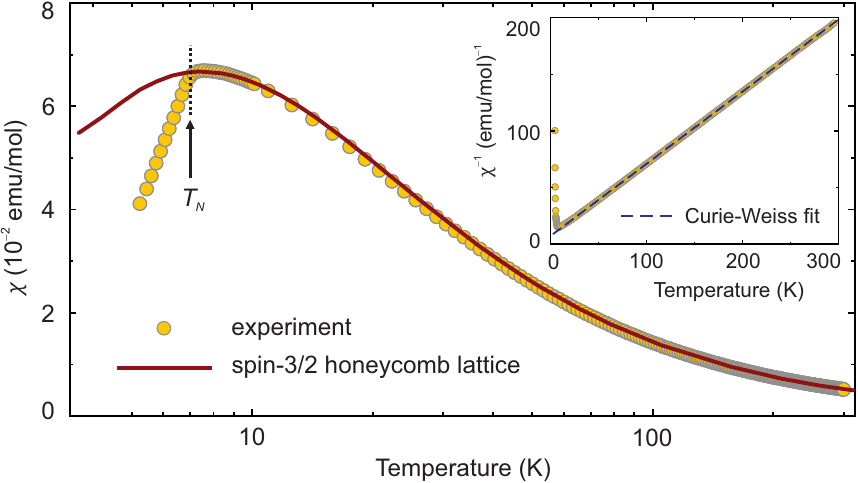}
\caption{Magnetic susceptibility ($\chi$) of MoP$_3$SiO$_{11}$ measured in the applied field of 0.01\,T (circles) and the fit with the model of spin-$\frac32$ honeycomb planes, $J=2.6$\,K (solid line). The inset shows inverse susceptibility and the Curie-Weiss fit.}
\label{fig:CHI}
\end{figure}


\subsection{Thermodynamic properties}
\label{sec:thermo}
Magnetic susceptibility of MoP$_3$SiO$_{11}$ reveals the Curie-Weiss behavior at high temperatures, followed by a peak around 7\,K (Fig.~\ref{fig:CHI}). Above 50\,K, the fit with the modified Curie-Weiss law, \mbox{$\chi(T) = \chi_0 + C/(T-\Theta)$}, returns the temperature-independent contribution $\chi_0 = (3.0 \pm 0.1) \times 10^{-5}$\,emu/mol, Curie constant $C=1.56\pm 0.06$\,emu\,K/mol, and Curie-Weiss temperature $\Theta = -10.7\pm 0.4$\,K. The Curie constant corresponds to the paramagnetic effective moment of 3.53\,$\mu_B$ and $g=1.82$ according to $\mu_{\rm eff}=g\sqrt{S(S+1)}$ for $S=\frac32$ of Mo$^{3+}$. The deviation from the free-electron value of $g\simeq 2.0$ gauges the effect of spin-orbit coupling on the single-ion physics. This effect is more pronounced than in Mo(PO$_3)_3$ with $g\simeq 1.93$~\cite{rojo2003} ($C=1.71$\,emu\,K/mol) and especially in the Cr$^{3+}$ compounds with $g\simeq 2.0$~\cite{janson2013,janson2014}. On the other hand, an even larger reduction in the $g$-value has been reported for Mo$^{4+}$ with $g=1.6-1.7$~\cite{lezama1995,hembacher2018}, because its $4d^2$ electronic configuration usually does not lead to quenching of the orbital moment.

\begin{figure}
\includegraphics{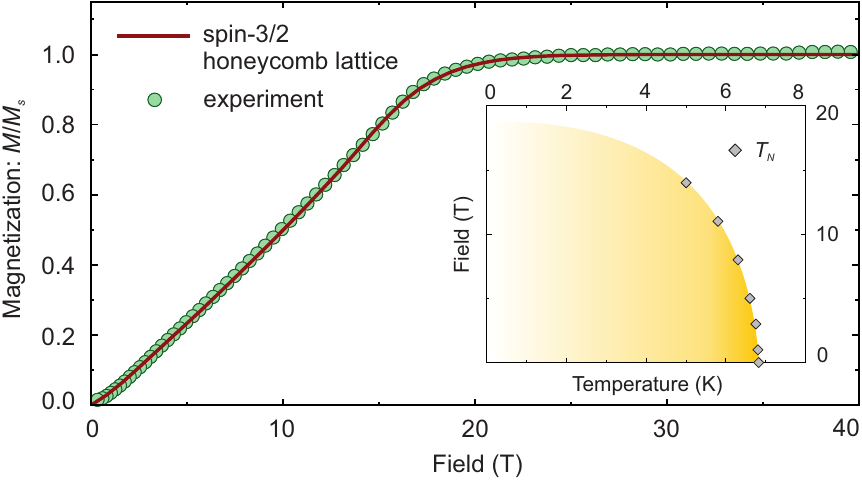}
\caption{Magnetization curve measured at 1.4\,K in pulsed fields, with absolute values of $M$ scaled to the saturated value $M_s$. The solid line is the fit with the model of spin-$\frac32$ honeycomb planes, $J=2.6$\,K. The inset shows field-temperature phase diagram, with diamonds depicting the $T_N$ values extracted from the specific-heat data (Fig.~\ref{fig:heat}). }
\label{fig:MAG}
\end{figure}

The negative value of $\Theta$ indicates predominant antiferromagnetic couplings that are independently gauged by the saturation field of about 18\,T observed in the pulsed-field measurement (Fig.~\ref{fig:MAG}). The susceptibility peak at $T_N=6.8$\,K reflects long-range antiferromagnetic ordering. This transition also manifests itself by the sharp $\lambda$-type anomaly in the specific heat. The ratio $\Theta / T_N \sim 1.6$ indicates a slight suppression of the long-range order that can be caused by a weak frustration or reduced dimensionality of the spin lattice. Our microscopic analysis (Sec.~\ref{sec:theory}) reveals the latter as the main cause. The entropy $S(T_N)$ released at $T_N$ corroborates the slight suppression of the long-range order. Indeed, by extrapolating $C_p(T)$ to $T\rightarrow 0$ and integrating $C_p/T$, we find $S(T_N)\simeq 5.0$\,J\,mol$^{-1}$\,K$^{-1}$, which is only 43\% of the total magnetic entropy $R\ln 4$ expected for $S=\frac32$.

\begin{figure}
\includegraphics{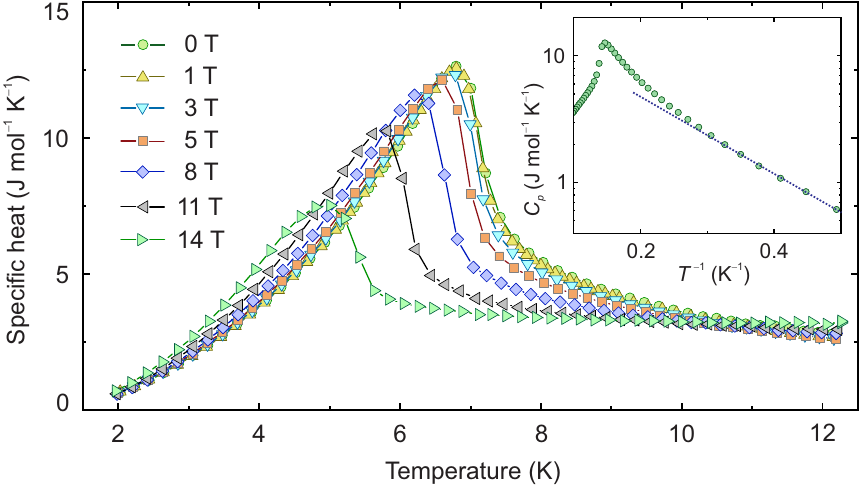}
\caption{Specific heat ($C_p$) of MoP$_3$SiO$_{11}$ measured in magnetic fields up to 14\,T. The inset shows activated behavior of the zero-field specific heat, with the dotted line used to determine the magnon gap $\Delta\simeq 7$\,K.}
\label{fig:heat}
\end{figure}

Below $T_N$, the specific heat deviates from the $T^3$ behavior, which would be expected in an isotropic antiferromagnet. The data below 3\,K follow the activated behavior, $C_p\sim e^{-\Delta/T}$, with the magnon gap $\Delta\simeq 7$\,K (Fig.~\ref{fig:heat}, inset). This exponential behavior of $C_p(T)$ indicates deviations from the simple Heisenberg model and a sizable magnetic anisotropy in MoP$_3$SiO$_{11}$. 

Specific heat measurements also reveal a gradual suppression of $T_N$ upon applying magnetic field (Fig.~\ref{fig:heat}). The resulting monotonic phase boundary in the $T-H$ phase diagram (Fig.~\ref{fig:MAG}, inset) distinguishes MoP$_3$SiO$_{11}$ from the low-dimensional spin-$\frac12$ antiferromagnets where $T_N$ first increases in low fields and then gets suppressed upon increasing the field further~\cite{tsyrulin2010,cizmar2010,tsirlin2011,tsirlin2013}. Such a non-monotonic behavior is rooted in the competition between quantum fluctuations of a Heisenberg antiferromagnet and uniaxial anisotropy introduced by the magnetic field~\cite{sengupta2009}. The monotonic phase boundary observed in MoP$_3$SiO$_{11}$ excludes this scenario and suggests the presence of magnetic anisotropy already in zero field.


\subsection{Magnetic structure}

\begin{figure}
\includegraphics{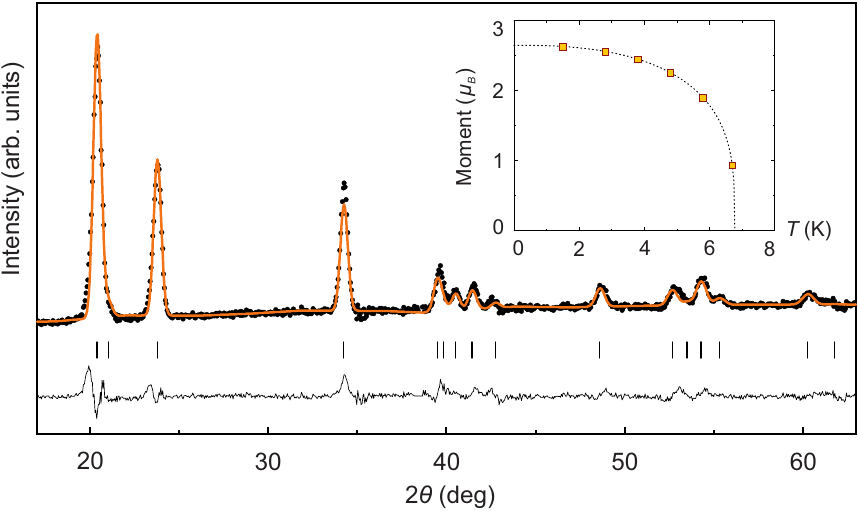}
\caption{Rietveld refinement for the magnetic neutron scattering obtained by subtracting the 12\,K data (above $T_N$) from the 1.5\,K data (below $T_N$). The orange line is the fit with the covalent magnetic form factor displayed in Fig.~\ref{fig:Form_factor}. The inset shows temperature dependence of the ordered magnetic moment and its empirical fit explained in the text. The error bars are smaller than the symbol size. }
\label{fig:Neutron}
\end{figure}

Neutron diffraction data give further insight into the magnetic transition in MoP$_3$SiO$_{11}$. No additional magnetic reflections were observed below $T_N$, but several low-angle reflections became more intense, suggesting $\kv=0$ as the propagation vector. Symmetry analysis based on the $R\bar 3c$ space group {\cred returns four irreps that are compatible with the fully compensated antiferromagnetic order, three of them are one-dimensional with the spins along $c$, whereas the fourth one is more complex and features spins in the $ab$ plane of the structure. Only this irrep led to a successful refinement (Fig.~\ref{fig:Neutron}). It comprises four basis vectors listed in Table~\ref{tab:irrep}. The magnetic structure is fully described by BV$_3$ that represents collinear spins pinned to the $[120]$ crystallographic direction (Fig.~\ref{fig:Model}) or one of the equivalent directions, $[210]$ and $[1\bar 10]$, obtained by a $60^{\circ}$ rotation. The corresponding} magnetic space group is $C2/c'$. 

\begin{table}
\caption{\label{tab:irrep}
Basis vectors of the irrep ($R\bar 3c$, $\kv=0$) that allows a fully compensated antiferromagnetic order with spins in the $ab$ plane. 
}
\begin{ruledtabular}
\begin{tabular}{lrrrr}
 Mo atom & BV$_1$ & BV$_2$ & BV$_3$ & BV$_4$ \\
 $(x,y,z)$                     & 1\,0\,0    & 0\,1\,0 & 0.5\,1\,0       & $\bar 1$\,$\overline{0.5}$\,0 \\
 $(x-y,\bar y,\bar z+\frac12)$ & $\bar 1$\,0\,0 & 1\,1\,0 & $\overline{0.5}$\,$\bar 1$\,0 & $\overline{0.5}$\,0.5\,0  \\
 $(\bar x,\bar y,\bar z)$ & $\bar 1$\,0\,0 & 0\,$\bar 1$\,0 & $\overline{0.5}$\,$\bar 1$\,0 & 1\,0.5\,0 \\
 $(-x+y,y,z+\frac12)$ & 1\,0\,0 & $\bar 1$\,$\bar 1$\,0 & 0.5\,1\,0 & 0.5\,$\overline{0.5}$\,0 \\
\end{tabular}
\end{ruledtabular}
\end{table}


\begin{figure}
\includegraphics[width=0.3\textwidth]{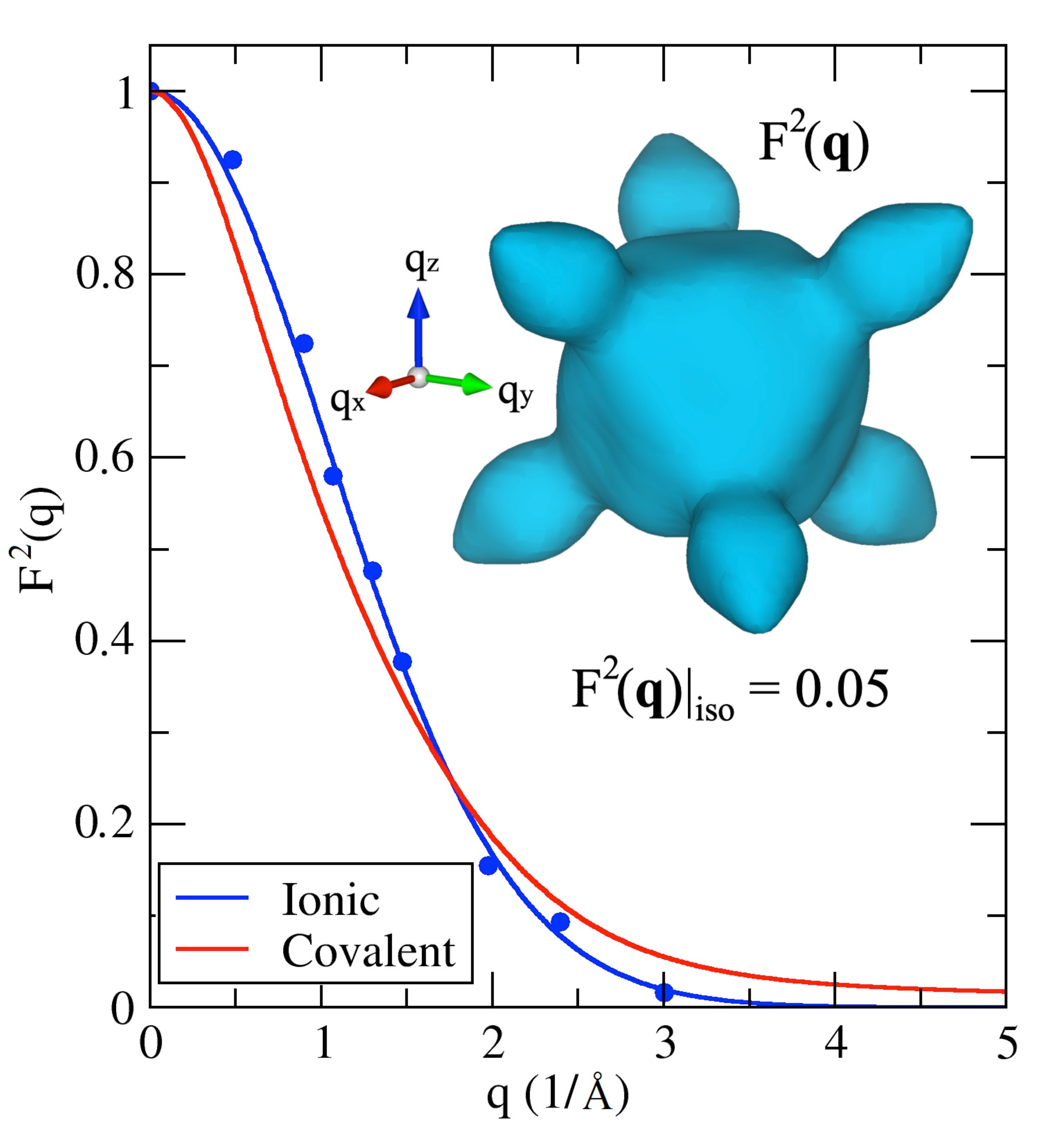}
\caption{
Powder-averaged momentum dependence of the magnetic form factor, $|F(q)|^2$, for Mo$^{3+}$. The standard parametrization~\citep{PJBrown} of $F(q)$ is given by $A=0.130$, $a=295.027$, $B=0.150$, $b=2.140$, $C=0.698$, $c=37.935$, and $D=0.022$ for the covalent form factor obtained in this work, and $A = 0.505$, $a = 43.558$, $B = 0.253$, $b = 28.016$, $C = 0.243$, $c = 28.804$, and $D = - 0.001$ for the ionic form factor (fit to the data from Ref.~\cite{wilkinson1961} shown as symbols). The inset shows the $|F(q)|^2=0.05$ isosurface for the covalent form factor.
}
\label{fig:Form_factor}
\end{figure}

Different approximations for the magnetic form factor of Mo$^{3+}$ were tested in the refinement. The ionic approximation was taken from Ref.~\cite{wilkinson1961} that reported the magnetic scattering from MoF$_3$. Alternatively, we compute the ``covalent'' form factor $F(\mathbf{q})$ by Fourier-transforming Wannier orbitals of $t_{2g}$ symmetry, $F(\mathbf{q}) = \int |W(\mathbf{r})|^2 e^{-i \mathbf{q r}}  d \mathbf{r}$~\cite{form_factor}. The orbitals are calculated for MoP$_3$SiO$_{11}$ and thus take all peculiarities of this compound into account. The resulting non-spherical shape of $F(\mathbf q)$ parallels the real-space arrangement of the three $t_{2g}$ orbitals (Fig.~\ref{fig:Form_factor}). Such a custom magnetic form factor shows a somewhat different $q$-dependence {\cred because the hybridization between the Mo $4d$ and O $2p$ orbitals (Mo--O covalency) is included. This} leads to a slight improvement of the refinement. Using the 1.5\,K data, we find $\mu=2.461(8)$\,$\mu_B$/Mo$^{3+}$ and $R=0.042$ with the ionic form factor, to be compared with $\mu=2.634(8)$\,$\mu_B$/Mo$^{3+}$ and $R=0.035$ when the covalent form factor is used. {\cred The 7\% difference in the refined magnetic moment indicates a marginal role of covalency in MoP$_3$SiO$_{11}$, to be compared with the Mo$^{4+}$-containing BaMoP$_2$O$_8$ where the 20\% difference has been reported, and a significant improvement in the quality of the refinement could be achieved~\cite{hembacher2018}.}

The temperature dependence of the ordered magnetic moment (Fig.~\ref{fig:Neutron}, inset) was fitted with the empirical function $\mu=\mu_0[1-(T/T_N)^{\alpha}]^{\beta}$ {\cred that can be used across a broad temperature range and should not be confused with the critical behavior (hence neither $\alpha$ nor $\beta$ are the true critical exponents)}. The fit returns $T_N=6.78$\,K, $\alpha=2.5$, $\beta=0.29$, and $\mu_0=2.65$\,$\mu_B$ as the zero-temperature value of the ordered magnetic moment. With $g=1.82$ from the Curie-Weiss fit, one expects $\mu_0=gS=2.73$\,$\mu_B$. The reduction from the spin-only value of 3\,$\mu_B$ is, thus, mostly caused by the spin-orbit coupling, while the reduction due to quantum fluctuations is minor.




\section{Microscopic modeling}
\label{sec:theory}

\subsection{Electronic structure}

\begin{figure}
\includegraphics[width=0.49\textwidth]{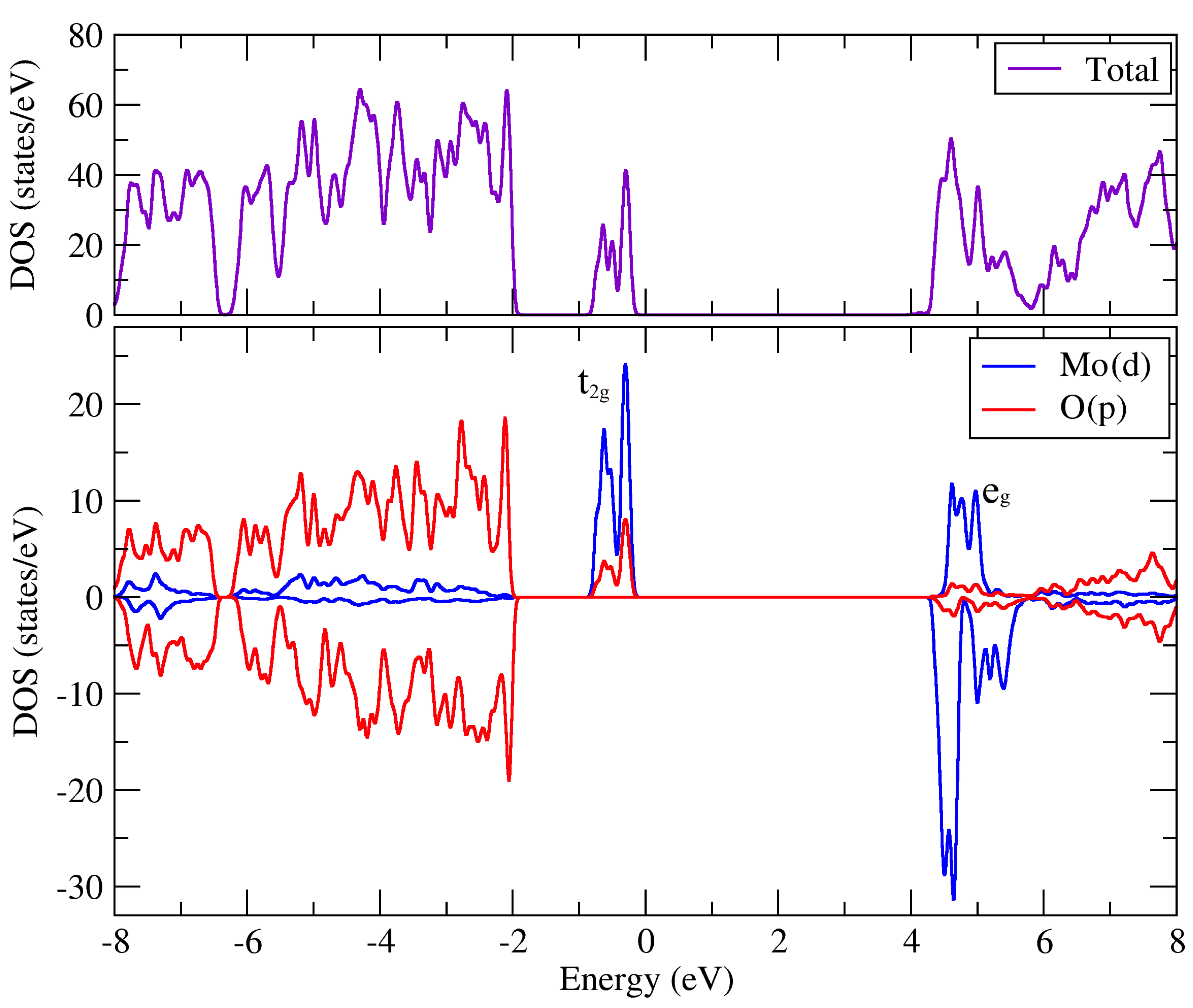}
\caption{{\cred (Top) GGA+$U$+SO total and (bottom) GGA+$U$ partial density of states (DOS) calculated for the ferromagnetic spin configuration of MoP$_3$SiO$_{11}$.} The corresponding atomic contributions are shown. The Fermi level is at zero energy. }
\label{fig:DOS}
\end{figure}

The electronic density of states (DOS) for MoP$_3$SiO$_{11}$ is shown in Fig.~\ref{fig:DOS}. {\cred One clearly identifies $t_{2g}$ as the magnetic orbitals, whereas the $e_g$ orbitals are empty, as expected for Mo$^{3+}$. The $t_{2g}$ states in the GGA band structure are} parametrized via maximally localized Wannier functions~\cite{marzari1997} using the local coordinate frame, where the axes $x^\prime y^\prime z^\prime$ are directed toward oxygen atoms of the MoO$_6$ octahedron. The resulting Wannier functions perfectly describe the GGA bands and yield nearest-neighbor hopping parameters within the honeycomb layer (Table~\ref{tab:Hopping}).

\begin{table}
\caption{\label{tab:Hopping}
Nearest-neighbor intralayer hopping parameters between the Mo$^{3+}$ ions (in meV). Magnetic sites are connected by the vectors $\mathbf{r}_{t_1} = (-4.21, -2.43, -0.60)$\,\AA, $\mathbf{r}_{t_2} = (4.21, -2.43, -0.60)$\,\AA, and $\mathbf{r}_{t_3} = (0.00, 4.86, -0.60)$\,\AA\ given relative to the magnetic unit cell with the $C2/c'$ symmetry.   
}
\begin{ruledtabular}
\begin{tabular}{ccc}
$\tensor{t}_{1}$ & $\tensor{t}_{2}$ &  $\tensor{t}_{3}$  \\
  \hline
$\left(\begin{array}{ccc} -16  & -18  & 43   \\ 43  & -35 & 25 \\    18 & -5  &  35   \end{array} \right)$  &  $\left(\begin{array}{ccc} 5  & -35  & -18   \\  -18  &  43 & -16 \\    35 & -25  &  43   \end{array} \right)$  &  $\left(\begin{array}{ccc} 25  & 43  & -35   \\  -35  & -18 &  5 \\    -43  &  16  &  18   \end{array} \right)$ 
\end{tabular}
\end{ruledtabular}
\end{table}

The hopping matrices $\tensor{t}_{1}$, $\tensor{t}_{2}$, and $\tensor{t}_{3}$ are transformed into each other by a
60$^\circ$ rotation about the $z$ axis, as expected for the regular, undistorted hexagonal lattice. Additionally, we find sizable hoppings along the shortest interlayer bond with the Mo--Mo distance of 7.264\,\r A,
\begin{equation}
\tensor{t}_c = \left(\begin{array}{ccc} -8  & -17  & 17   \\ 17  & -8 & 17 \\    17 &  17  &  -8   \end{array} \right)
\end{equation} 
where the values are given in meV. Similar to the in-plane couplings, all three $t_{2g}$ orbitals contribute to the hoppings. Other interlayer interactions are weak because they feature hopping matrices with a significant contribution from one orbital only. The selection of $J_c$ as the dominant interlayer coupling is probably caused by the triple P$_2$O$_7$ bridges that are present for this coupling but absent for any other interlayer Mo--Mo contact (Fig.~\ref{fig:Crystal}b). The experimental magnetic structure (Fig.~\ref{fig:Anisotropy}) reveals antiparallel spin arrangement along the $J_c$ pathway and confirms $J_c$ as the leading interlayer coupling.


\subsection{Exchange interactions}
We will now compute relevant magnetic interactions in MoP$_3$SiO$_{11}$. To this end, we define the spin Hamiltonian 
\begin{eqnarray}
\mathcal {\hat{H}} =  J  \sum\limits_{i > j}^{nn}  \hat{\mathbf{S}}_i \hat{\mathbf{S}}_j + J_c  \sum\limits_{i > j}^{nnn}  \hat{\mathbf{S}}_i \hat{\mathbf{S}}_j  + D \sum\limits_{i } \hat{S}^2_{iz}   
\label{eq:Magnetic_model}
\end{eqnarray} 
with isotropic couplings within ($J$) and between ($J_c$) the honeycomb planes. Magnetic anisotropy is introduced by the single-ion term $D$, which is defined relative to the three-fold axis of the crystal structure. In Sec.~\ref{sec:anisotropy}, we show that magnetic anisotropy energy of MoP$_3$SiO$_{11}$ is dominated by this term, whereas contributions of intersite terms are negligible. 

\begin{figure}
\includegraphics{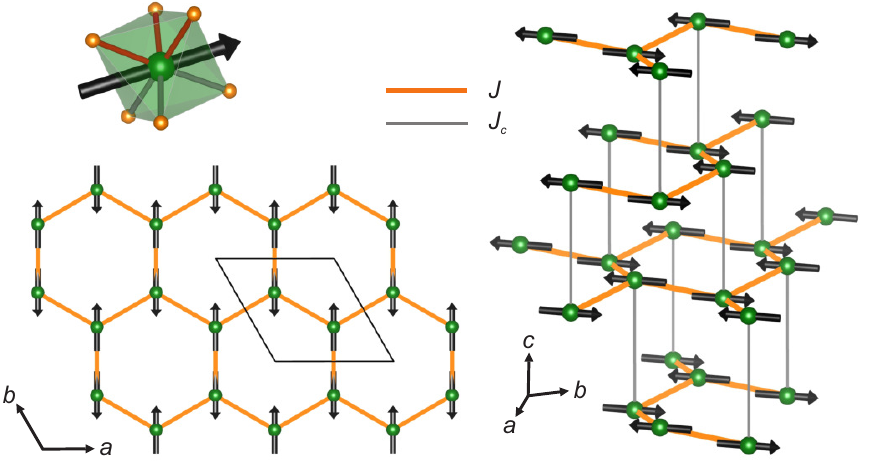}
\caption{Experimental magnetic structure of MoP$_3$SiO$_{11}$ superimposed on the spin lattice with the in-plane coupling $J$ and interplane coupling $J_c$. }
\label{fig:Model}
\end{figure}

The respective spin lattice is visualized in Fig.~\ref{fig:Model}. In contrast to other hexagonal magnets, each lattice site features only one interlayer coupling, either to the layer above or to the layer below. This unusual coupling scheme is a corollary of the $ABCABC$ stacking sequence that allows only one $J_c$ contact per Mo$^{3+}$ ion (Fig.~\ref{fig:Crystal}b). 

Exchange interactions are calculated by a mapping procedure~\cite{xiang2011} using total energies of collinear spin configurations,
\begin{eqnarray}
J_{ij} =  \frac{1}{4zS^2} (E_{\uparrow \uparrow} + E_{\downarrow \downarrow} - E_{\uparrow \downarrow} - E_{\downarrow \uparrow}),
\end{eqnarray}
where $z$ is the number of neighbors, which have the same interaction $J_{ij}$. $E_{\uparrow \downarrow}$ represents the total energy of the spin state with opposite directions of the magnetic moments on the sites $i$ and $j$. Total energies are calculated on the DFT+$U$+SO level with the Hund's coupling $J_H=0.8$\,eV and on-site Coulomb repulsion $U=4-5$\,eV, which is higher than the optimal $U=3$\,eV for Mo$^{4+}$ in BaMoP$_2$O$_8$~\cite{hembacher2018} because the lower oxidation state of Mo in MoP$_3$SiO$_{11}$ reduces covalency and screening. The double-counting correction in the atomic limit was used~\cite{double_counting}.

Exchange interactions obtained for several values of the $U$ parameter are listed in Table~\ref{tab:Exchange_couplings}. Both $J$ and $J_c$ decrease upon increasing $U$, as typical for the kinetic antiferromagnetic superexchange arising from the electron hoppings. 

\begin{table}
\caption{\label{tab:Exchange_couplings}
Isotropic exchange interactions $J$ and $J_c$, magnetic moment $\mu$ of the Mo$^{3+}$ ion, and the Curie-Weiss temperature $\Theta$ depending on the on-site Coulomb repulsion parameter $U$ within DFT+$U$+SO.  
}
\begin{ruledtabular}
\begin{tabular}{c|ccc}
   & $U$ = 4\,eV & $U$ = 4.5\,eV & $U$ = 5\,eV    \\
  \hline
 $J$ (K)      & 2.80 &  2.60 & 2.15 \\
 $J_c$ (K) & 0.90 &  0.80 & 0.70 \\  
 $\mu$ $(\mu_B)$  & 2.70  & 2.72 & 2.75  \\
 $\Theta$  (K) & $-11.6$ &  $-10.8$ &  $-8.9$ \\
 \end{tabular}
\end{ruledtabular}
\end{table}

The optimal value of $U$ is chosen on the basis of the Curie-Weiss temperature calculated as 
\begin{eqnarray}
\Theta = -\frac{S(S+1)}{3} (3 J + J_c)
\end{eqnarray}
and compared to the experimental value of $-10.7\pm 0.4$\,K that returns $U=4.5$\,eV. In the following, we take the same value of $U$ to calculate the magnetic anisotropy of the Mo$^{3+}$ ion.


\subsection{Magnetic anisotropy}
\label{sec:anisotropy}
\begin{figure}
\includegraphics[width=0.49\textwidth]{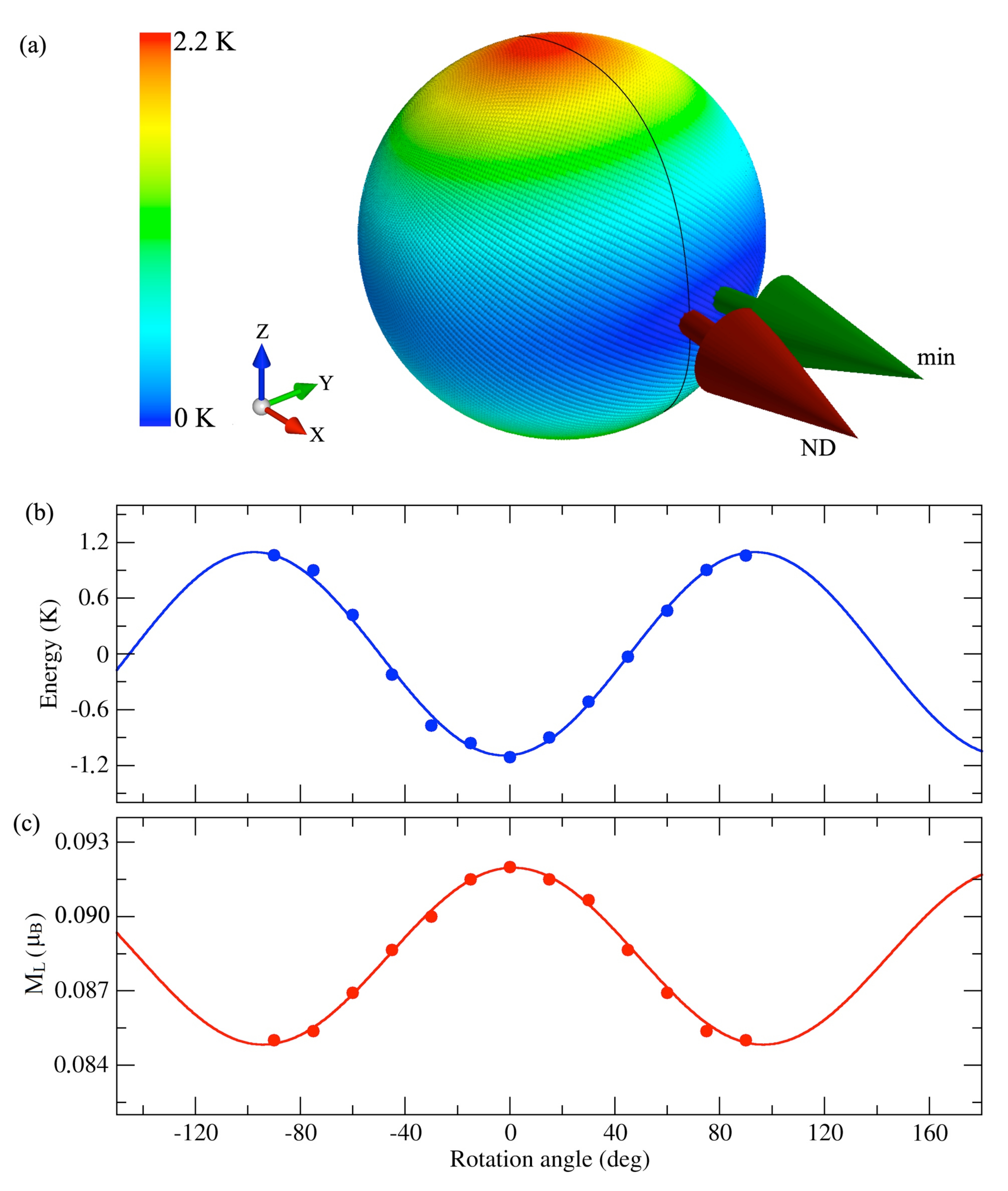}
\caption{(a) Magnetic anisotropy energy $E_{\rm anis}/S^2$ obtained by extrapolating a finite number of spin configurations. The red arrow "ND" denotes the magnetic moment direction from the neutron diffraction experiment, whereas the green arrow "min" stands for the optimal direction in DFT+$U$+SO. {\cred This optimal direction is stabilized by the tiny anisotropy within the easy plane.} The black line schematically represents the rotation plane for (b-c). (b) The dependence of $E/S^2$ on the rotation of a single spin in the plane spanned by $[120]$ ($\varphi=0^{\circ}$, energy minimum) and $[001]$ ($\varphi=90^{\circ}$, energy maximum). (c) Angular dependence of the orbital moment upon the same rotation. 
}
\label{fig:Anisotropy}
\end{figure}

We first analyze changes in the total energy upon a uniform rotation of all spins. Using a finite number of representative spin directions~\cite{mazurenko_Mn12}, we extrapolate the energy dependence for an arbitrary direction over a sphere surrounding the magnetic ion (Fig.~\ref{fig:Anisotropy}a). The energy distribution gives a clear witness of the easy-plane anisotropy and identifies $[001]$ as the magnetic hard axis with the respective anisotropy energy of $E_{\rm anis}\simeq 5.0$\,K. Additionally, a weak anisotropy of about 0.1\,K is found in the $ab$ plane {\cred beyond the anisotropy term included in Eq.~\eqref{eq:Magnetic_model}}. The lowest energy is obtained for spins directed along $[120]$ in agreement with the magnetic moment direction determined experimentally.

{\cred In the following, we focus on the leading easy-plane anisotropy that distinguishes the $ab$ plane from $c$ as the magnetic hard axis.} This magnetic anisotropy can arise from both single-ion and intersite terms. They are distinguished in a calculation where one spin is rotated in the plane spanned by $[120]$ {\cred (a direction within the easy plane)} and $[001]$ {\cred (hard axis)}, while all neighboring spins are orthogonal to this plane. This procedure eliminates the contribution of intersite anisotropy and separates the single-ion term $D\simeq 2.2$\,K (Fig.~\ref{fig:Anisotropy}b). It gives the major contribution to the anisotropy energy, $DS^2\simeq 4.95$\,K, and follows the angular dependence of the orbital moment~\cite{bloch1931, bruno1989} (Fig.~\ref{fig:Anisotropy}c). The highest value of the orbital moment is obtained for the direction within the easy plane. This orbital moment is antiparallel to the spin moment, thus supporting the reduction of the $g$-value below 2.0. The total moment of $3-0.09\simeq 2.9$\,$\mu_B$ is slightly higher than 2.73\,$\mu_B$ expected from $g=1.82$ determined experimentally.


\subsection{Model simulations}
\label{eq:simulations}
We will now compare these microscopic results with the experimental data. We first use the simplest magnetic model of decoupled honeycomb planes ($J_c=0$, $D=0$) that allows a decent description of the high-field magnetization data (Fig.~\ref{fig:MAG}), as well as of the magnetic susceptibility data down to $T_N$ (Fig.~\ref{fig:CHI}). The fitted parameters of $J=2.6$\,K and $g=1.82$ are in an excellent agreement with the results of our \textit{ab initio} calculations and Curie-Weiss analysis, respectively.

One aspect missing in this simplified description is the formation of the long-range magnetic order that would be forbidden at any finite temperature in the 2D Heisenberg antiferromagnet with $J_c=0$ and $D=0$. Either of $J_c$ and $D$ leads to a finite $T_N$ that we evaluated using scaling properties of the spin stiffness~\cite{sengupta2009,tsirlin2012} obtained from QMC. With $D=2.2$\,K ($D/J=0.85$), one expects $T_N=5.2$\,K, which is slightly below the experimental value of 6.8\,K in zero field. By the same token, the interlayer coupling $J_c=0.8$\,K alone ($J_c/J=0.3$) would yield the underestimated value of $T_N=5.5$\,K. Using both $D/J=0.85$ and $J_c/J=0.3$, we arrive at $T_N=7.1$\,K in an excellent agreement with the experiment, thus completing the microscopic description of MoP$_3$SiO$_{11}$. Moreover, from the spin-wave expression for the magnon gap $\Delta=2DS=3D$~\cite{aguilera2020} we estimate $\Delta=6.6$\,K in a remarkable agreement with the experimental value of about 7\,K (Sec.~\ref{sec:thermo}). {\cred Note that this gap is associated with the easy-plane anisotropy of MoP$_3$SiO$_{11}$. The additional weak anisotropy within the easy plane could not be resolved in our present experiments and requires a further dedicated study.}

It is also worth noting that our experimental value of the ordered magnetic moment is only marginally reduced compared to 3\,$\mu_B$ expected for spin-$\frac32$. This reduction is mostly accounted for by the weak orbital moment inferred from $g=1.82$. Therefore, quantum fluctuations are expected to play a minor role in the ground state. Indeed, estimating the ordered magnetic moment for the $J-J_c$ model ($D=0$) via the standard extrapolation procedure~\cite{Sandvik_finite_size, Danis_Ba3Cu3Sc4O12} results in 2.72\,$\mu_B$, about 10\% reduction. This can be compared to the 45\% reduction in the honeycomb antiferromagnet with spin-$\frac12$~\cite{Sandvik_honeycomb}. Large single-ion anisotropy should suppress quantum fluctuations even further.

\section{Discussion and Summary}
\label{sec:discussion}

MoP$_3$SiO$_{11}$ is a $4d^3$ honeycomb antiferromagnet with the disconnected MoO$_6$ octahedra. It reveals a sizable spatial anisotropy of exchange couplings ($J_c/J = 0.35$) but also a substantial single-ion magnetic anisotropy $D$, which is similar in magnitude to the leading exchange coupling $J$. This anisotropy chooses the in-plane spin direction of the collinear antiferromagnetic order established by $J$ and $J_c$. It also enhances $T_N$ and keeps the ordered magnetic moment close to its classical value. The size of the ordered moment in MoP$_3$SiO$_{11}$ can be accounted for by the weak orbital contribution without invoking quantum effects. This indicates only a minor role of quantum fluctuations in the ground state of this spin-$\frac32$ antiferromagnet.

Other $4d^3$ honeycomb antiferromagnets show a very different balance between $D$ and $J$. For example, SrRu$_2$O$_6$ features $|D|/J=0.028$~\cite{suzuki2019} to be compared with $D/J = 0.85$ in MoP$_3$SiO$_{11}$. This leads to a significant difference in the relative size of the magnon gap, $\Delta/J=0.83$ in SrRu$_2$O$_6$ vs. $\Delta/J=2.7$ in MoP$_3$SiO$_{11}$. Both differences can be traced back to the reduction in $J$ for MoP$_3$SiO$_{11}$ with its disconnected MoO$_6$ octahedra. However, also the $D$ value changes drastically, from $-14$\,K (easy-axis anisotropy) in SrRu$_2$O$_6$ to $+2.2$\,K (easy-plane anisotropy) in MoP$_3$SiO$_{11}$. This change correlates with the local distortion of the transition-metal octahedra that show the compression along the three-fold axis in SrRu$_2$O$_6$ ($\alpha=93.2^{\circ}$) vs. elongation along the three-fold axis in MoP$_3$SiO$_{11}$ ($\alpha=88.8^{\circ}$), see Fig.~\ref{fig:Crystal} for the definition of the angle $\alpha$. In both cases, single-ion anisotropy has a strong impact on the magnetism, despite the quenched orbital moment of the $d^3$ ions. Unquenching the orbital moment increases the anisotropy further. For example, one finds $|D|\simeq 70$\,K in BaMoP$_2$O$_8$ with Mo$^{4+}$~\cite{abdeldaim2019}.

In summary, we revised the crystallographic symmetry of the MoP$_3$SiO$_{11}$ silicophosphate and showed that its space group is $R\bar 3c$, resulting in the formation of perfect honeycomb planes of the spin-$\frac32$ Mo$^{3+}$ ions. {\cred Their magnetic moments are quite robust as a result of the half-filled $t_{2g}$ shell and Hund's coupling.} Magnetic couplings within the honeycomb planes ($J\simeq 2.6$\,K) are three times stronger than the interplane couplings ($J_c\simeq 0.8$\,K). Collinear antiferromagnetic order caused by $J$ and $J_c$ is reinforced by the sizable easy-plane anisotropy $D\simeq 2.2$\,K that opens the magnon gap $\Delta\simeq 7$\,K. Our data suggest single-ion anisotropy as a major ingredient of $4d^3$ magnets despite their nominally quenched orbital moment. Both sign and size of this anisotropy may be controlled by local deformations of the transition-metal octahedra.

\acknowledgments
We are grateful to Anton Jesche for his continuous lab support and essential advice on thermodynamic measurements. The beam time provided by the ILL~\cite{neutron} and ESRF~\cite{esrf} was instrumental for completing this project. We thank Andy Fitch, Catherine Dejoie, and Vera Bader for their help during the data collection at ID22. We also acknowledge support by the HLD at HZDR, member of the European Magnetic Field Laboratory (EMFL). 

The work of D.I.B. was supported by the Russian Science Foundation, Grant No. 21-72-10136. The work in Augsburg was supported by the Federal Ministry for Education and Research through the Sofja Kovalevskaya Award of Alexander von Humboldt Foundation (A.A.T) and by the German Research Foundation (DFG) via the Project No. 107745057 (TRR80). L.D. acknowledges support from the Rutherford International Fellowship Programme (RIFP). This project has received funding from the European Union’s Horizon 2020 research and innovation
program under Marie Sk{\l}odowska-Curie Grant Agreement No. 665593 awarded to the Science and Technology Facilities Council.


%

\end{document}